\documentclass[preprint,nofootinbib,floats,tightenlines,amsfonts,amsmath,amssymb]{revtex4}

\usepackage{graphicx}
\usepackage{bm}
\usepackage{color}
\usepackage{multirow}
\usepackage{slashed}

\begin{document}
\baselineskip=17pt \parskip=5pt

\preprint{NCTS-PH/1901}
\hspace*{\fill}

\title{Rare hyperon decays with missing energy}

\author{Jusak Tandean}
\email{jtandean@yahoo.com}
\affiliation{Department of Physics, National Taiwan University,
No.\,\,1, Sec.\,\,4, Roosevelt Rd., Taipei 106, Taiwan \smallskip \\
\mbox{Physics Division, National Center for Theoretical Sciences},
No.\,\,101, Sec.\,\,2, Kuang Fu Rd., Hsinchu 300, Taiwan
\bigskip}


\begin{abstract}

We explore the strangeness-changing decays of the lightest hyperons into another baryon plus
missing energy within and beyond the standard model (SM).
In the SM these processes arise from the loop-induced quark transition $s\to d\nu\bar\nu$ and
their branching fractions are estimated to be less than $10^{-11}$.
In the presence of new physics (NP) the rates of these hyperon decays with missing energy could
increase significantly with respect to the SM expectations because of modifications to the SM
process or contributions from additional modes with new invisible particles.
Adopting a model-independent approach and taking into account constraints from the kaon sector,
we find that the current data on $K\to\pi\nu\bar\nu$ do not permit sizable NP impact on
the hyperon decays via underlying operators having mainly parity-even quark parts.
In contrast, NP operators with primarily parity-odd quark parts are much less
restricted by the existing bounds on $K\to\rm invisible$ and $K\to\pi\pi\nu\bar\nu$ and
consequently could produce substantially amplifying effects on the hyperon modes.
Their NP-enhanced branching fractions could reach levels potentially observable in the ongoing
BESIII experiment.

\end{abstract}

\maketitle

\section{Introduction \label{intro}}

The strangeness-changing quark transition \,$s\to d\slashed E$,\, with missing energy $\slashed E$
in the final state, is of great interest because it serves as an environment in which to test
the standard model (SM) and therefore also to look for signals of possible new physics (NP)
beyond it.
Within the SM, this process is predominantly due to \,$s\to d\nu\bar\nu$\, arising from
$Z$-penguin and box diagrams with up-type quarks and the $W$ boson in
the loops \cite{Buchalla:1995vs}, the neutrino pair ($\nu\bar\nu$) being undetected.
In the presence of NP, this SM contribution could be altered \cite{Marciano:1996wy,
Grossman:1997sk,Buras:2004uu,Lee:2015qra,Fuyuto:2015gmk,Li:2016tlt,Chiang:2017hlj,
Hu:2018luj},
and there might be invisible nonstandard states which are light enough and have $sd$ couplings
to give rise to new \,$s\to d\slashed E$\, channels \cite{He:2004it,Bird:2006jd,He:2010nt,
Kamenik:2011vy,Dolan:2014ska,Gninenko:2015mea,Abada:2016plb,
He:2018uey,Barducci:2018rlx,Matsumoto:2018acr}.

Currently there are ongoing efforts to observe \,$s\to d\nu\bar\nu$\, via the kaon decays
\,$K^+\to\pi^+\nu\bar\nu$\, and \,$K_L\to\pi^0\nu\bar\nu$\, by the NA62 \cite{CortinaGil:2018fkc}
and KOTO \cite{Ahn:2018mvc} Collaborations, respectively.
These measurements might then probe for hints of \,$s\to d\slashed E$\, beyond the SM as well.
Additional kaon modes worth pursuing are \,$K_L\to\slashed E$\, and \,$K\to\pi\pi'\slashed E$,\,
as only moderate bounds on \,$K\to\pi\pi'\nu\bar\nu$\, from direct searches are
available~\cite{Tanabashi:2018oca}.
Hence improved data on these extra modes would also be desirable.
In the baryon sector, their counterparts are the strangeness-changing ($|\Delta S|=1$) decays of
light hyperons into another baryon plus missing energy, on which there is still no empirical
information.
Interestingly, measuring such processes in the BESIII experiment has recently been proposed
and may be realized in the near future \cite{Li:2016tlt}.

Here we have a look at these rare hyperon decays to investigate how much they may be affected
by different NP possibilities, taking into account restrictions from the kaon sector.
Initial studies on the hyperon modes due to new \,$s\to d\nu\bar\nu$\, interactions with the same
chiral structure as in the SM have been carried out in refs.\,\cite{Li:2016tlt,Hu:2018luj}.
In this paper, we explore a more general scenario in which the underlying NP operators
might involve other Lorentz structures and the invisible pair could be nonstandard fermions.
It turns out that in this more general case the hyperon rates may be significantly enlarged with
respect to their SM values and even reach potentially discoverable levels at BESIII.

As detailed later on in this analysis, such a possibility has to do with the fact that
these kaon and hyperon decays do not probe the same portions of the underlying
\,$s\to d\slashed E$\, operators and with the kaon data situation at the moment.
Particularly, \,$K\to\pi\slashed E$\, is sensitive exclusively to the terms in the operators which
have parity-even quark parts and \,$K\to\slashed E$\, to the terms having parity-odd quark parts,
whereas \,$K\to\pi\pi'\slashed E$\, and the hyperon modes can probe both.
Given that the latest measurements of $K\to\pi\nu\bar\nu$ decays
\cite{CortinaGil:2018fkc,Ahn:2018mvc,Tanabashi:2018oca} have left only little room for NP to
influence them, it follows that NP cannot raise the hyperon rates considerably
above their SM expectations if it enters via $s\to d\slashed E$\, operators with
mainly parity-even quark parts.
On the other hand, the constraints from the existing data on \,$K\to\slashed E$ and
\,$K\to\pi\pi'\slashed E$\, are relatively much weaker, implying that NP operators having primarily
parity-odd quark parts are still allowed to yield sizable enhancing effects on the hyperon modes.

The organization of the paper is as follows.
In section \ref{Lf} we write down a number of effective low-energy operators contributing to
\,$s\to d\slashed E$\, which may be generated by NP.
Without getting into model specifics, we treat the operators in a model-independent manner.
In section \ref{hyperons} we first deal with the baryonic matrix elements pertinent to
the corresponding hyperon decays with missing energy and subsequently derive their
differential rates.
Similarly, in section \ref{kaons} we provide the formulas for the rare kaon decays
of concern.
In section \ref{numeric} we present our numerical results.
We begin by evaluating the SM predictions for the hyperon modes and comparing them to
the proposed sensitivity reach of BESIII.
Next we look at the kaon sector and examine its restraints on NP impacting the operators.
We then show that some of the allowed NP couplings can amplify the hyperon rates to values
that may be observable by BESIII.
In section \ref{concl}, we give our conclusions.
We collect extra formulas and further details in a couple of appendixes.

\section{Interactions\label{Lf}}

Beyond the SM, there could be new ingredients which induce modifications to the \,$s\to d\nu\bar\nu$
transition in the SM and/or bring about additional \,$s\to d\slashed E$\, channels with one or more
invisible light nonstandard bosons or fermions emerging in the final states.
These new particles could be stable or sufficiently long-lived to escape detection.
Among such possibilities, in this study we focus on the \,$s\to d\slashed E$\, scenario in which
the missing energy is due to \,$\texttt{\textsl f}\bar{\texttt{\textsl f}}$\, being emitted where
$\texttt{\textsl f}$ is an electrically neutral, uncolored, and invisible Dirac fermion having
spin 1/2.
Thus, $\texttt{\textsl f}$ could be the SM neutrino $\nu$, which is not detected,
or a nonstandard fermion.

We consider $sd_{\;}\!\texttt{\textsl f}\bar{\texttt{\textsl f}}$ interactions described by
the low-energy effective Lagrangian
\begin{align} \label{Lnp}
{\cal L}_{\texttt{\textsl f}}^{} & \,=\, -\Big[ \overline{d}\gamma^\eta s~
\overline{\texttt{\textsl f}} \gamma_\eta^{} \big( \texttt C_{\texttt{\textsl f}}^{\texttt V}
+ \gamma_5^{} \texttt C_{\texttt{\textsl f}}^{\texttt A} \big) \texttt{\textsl f}
+ \overline{d} \gamma^\eta\gamma_5^{} s~ \overline{\texttt{\textsl f}} \gamma_\eta^{}
\big( \tilde{\textsf c}_{\texttt{\textsl f}}^{\textsc v}
+ \gamma_5^{} \tilde{\textsf c}{}_{\texttt{\textsl f}}^{\textsc a} \big) \texttt{\textsl f}
\nonumber \\ & ~~~~ ~~~~ +\,
\overline{d} s~ \overline{\texttt{\textsl f}} \big( \texttt C_{\texttt{\textsl f}}^{\texttt S}
+ \gamma_5^{} \texttt C_{\texttt{\textsl f}}^{\texttt P} \big) \texttt{\textsl f}
+ \overline{d} \gamma_5^{}s~ \overline{\texttt{\textsl f}}
\big( \tilde{\textsf c}_{\texttt{\textsl f}}^{\textsc s} + \gamma_5^{}
\tilde{\textsf c}_{\texttt{\textsl f}}^{\textsc p} \big) \texttt{\textsl f} \Big]
+\, {\rm H.c.} \,, &
\end{align}
where in our model-independent approach
$\texttt C_{\texttt{\textsl f}}^{\texttt V,\texttt A,\texttt S,\texttt P}$ and
$\tilde{\textsf c}_{\texttt{\textsl f}}^{\textsc v,\textsc a,\textsc s,\textsc p}$ are free
parameters which are generally complex and have the dimension of inverse squared mass.
The terms in ${\cal L}_{\texttt{\textsl f}}$ are Lorentz invariant and respect the unbroken
SU(3)$_{\rm color}\times$U(1)$_{\textsc{em}}$ gauge symmetry.\footnote{For $\texttt{\textsl f}$
being a particle from a dark sector beyond the SM, the terms in ${\cal L}_{\texttt{\textsl f}}$
would constitute a subset of the independent operators detailed in \cite{Kamenik:2011vy},
which include those containing dark particles of spin 0, 1, or 3/2.}
The grouping of the different operators in ${\cal L}_{\texttt{\textsl f}}$ according to
the parity of their quark bilinears is convenient because in the decay rates to be examined
the contributions of $\texttt C_{\texttt{\textsl f}}^{\texttt V,\texttt A,\texttt S,\texttt P}$,
belonging to ${\cal L}_{\texttt{\textsl f}}$ terms with parity-even quark bilinears, do not
interfere with the contributions of
$\tilde{\textsf c}_{\texttt{\textsl f}}^{\textsc v,\textsc a,\textsc s,\textsc p}$,
belonging to the terms with parity-odd quark bilinears.

As we will concentrate on exploring the potential implications of NP encoded in
${\cal L}_{\texttt{\textsl f}}$ for the transitions of light baryons and mesons at low hadronic
energies, here we will not concern ourselves with how the parameters in
${\cal L}_{\texttt{\textsl f}}$ evolve from their high-energy values.
It is nevertheless worth mentioning that for effective flavor-changing operators involving light
quarks and SM-gauge-singlet dark particles the running of their coefficients from high to low
energies has been estimated to be negligible~\cite{Arteaga:2018cmw}.

In what follows we address how the NP may enlarge the rates of \,$|\Delta S|=1$\, hyperon decays
with missing energy compared to the SM expectations.
Since ${\cal L}_{\texttt{\textsl f}}$ influences kaon decays as well, we need to ensure that
the applied ranges of $\texttt C_{\texttt{\textsl f}}^{\texttt V,\texttt A,\texttt S,\texttt P}$
and $\tilde{\textsf c}_{\texttt{\textsl f}}^{\textsc v,\textsc a,\textsc s,\textsc p}$ are
compatible with the available relevant data.
In numerical work, we will assume the phenomenological viewpoint that these free parameters can
have any values consistent with the empirical restrictions and perturbativity,
and so some of them may be taken to be vanishing or much smaller than the others.
This will allow us to look for parameter ranges that would translate into the maximal hyperon
rates permitted by the kaon constraints.
Although this may entail substantial differences among the NP parameters that appear rather
unnatural, we optimistically suppose that models could be devised to accommodate
them.\footnote{For instance, a model possessing a heavy $Z'$ boson which has family-nonuniversal
purely-vector interactions with SM quarks \cite{Langacker:2008yv} might be responsible for
the $\overline{d}\gamma^\eta s$ terms in ${\cal L}_{\texttt{\textsl f}}$. If the $Z'$ couplings
to SM quarks are purely axial-vector instead \cite{Langacker:2008yv,Ismail:2016tod,Alves:2016cqf},
it might give rise to the $\overline{d}\gamma^\eta\gamma_5^{}s$ terms.\medskip}

\section{Baryon decays\label{hyperons}}

Our hyperon decays of interest are
\,$\mathfrak B\to\mathfrak B'\texttt{\textsl f}\bar{\texttt{\textsl f}}$\, with
\,$\mathfrak{BB}'=\Lambda n,\Sigma^+p,\Xi^0\Lambda,\Xi^0\Sigma^0,\Xi^-\Sigma^-$,\,
all the particles being spin-1/2 fermions,\footnote{We do not include
\,$\Sigma^0\to n\texttt{\textsl f}\bar{\texttt{\textsl f}}$\, because its branching fraction
is expected to be comparatively very suppressed due to the $\Sigma^0$ width being overwhelmingly
dominated by the electromagnetic channel \,$\Sigma^0\to\Lambda\gamma$ \cite{Tanabashi:2018oca}.}
and \,$\Omega^-\to\Xi^-\texttt{\textsl f}\bar{\texttt{\textsl f}}$,\, where $\Omega^-$ is
a spin-3/2 hyperon.
Some of these modes may be looked for in the near future at BESIII where, with one year's
integrated luminosity, as many as $10^6$-$10^8$ hyperons ($\Lambda$, $\Sigma$, $\Xi$, and $\Omega$)
may be produced in $J/\psi$ and $\psi(2S)$ decays \cite{Li:2016tlt}.

To calculate the amplitudes for the hyperon decays, we need to know the baryonic matrix elements
of the quark parts of the operators in eq.\,(\ref{Lnp}).
We estimate the matrix elements with the aid of flavor-SU(3) chiral perturbation theory at
leading order.
Their derivation from the leading-order chiral Lagrangian is
outlined in appendix \ref{quark-hadron}.
We express the results pertaining to
\,$\mathfrak B\to\mathfrak B'\texttt{\textsl f}\bar{\texttt{\textsl f}}$\, as
\begin{align} \label{<B'B>}
\big\langle\mathfrak B'\big|\overline{d}\gamma^\eta s\big|\mathfrak B\big\rangle & \,=\,
{\cal V}_{\mathfrak B'\mathfrak B}^{}\,\bar u_{\mathfrak B'}^{}\gamma^\eta u_{\mathfrak B}^{} \,,
~ & \big\langle{\mathfrak B'}\big|\overline{d}\gamma^\eta\gamma_5^{}s\big|{\mathfrak B}\big\rangle
& \,=\, \bar u_{\mathfrak B'}^{} \bigg( \gamma^\eta {\cal A}_{\mathfrak B'\mathfrak B}^{}
- \frac{{\cal P}_{\mathfrak B'\mathfrak B}}{B_0}\, \texttt Q^\eta{} \bigg)
\gamma_5^{} u_{\mathfrak B}^{} \,,
\nonumber \\
\big\langle{\mathfrak B}'\big|\overline{d}s\big|{\mathfrak B}\big\rangle & \,=\,
{\cal S}_{\mathfrak B'\mathfrak B}^{}\, \bar u_{\mathfrak B'}^{} u_{\mathfrak B}^{} \,, &
\big\langle{\mathfrak B'}\big|\overline{d} \gamma_5^{}s\big|{\mathfrak B}\big\rangle
& \,=\, {\cal P}_{\mathfrak B'\mathfrak B}^{}\,
\bar u_{\mathfrak B'}^{}\gamma_5^{}u_{\mathfrak B}^{} \,,
\end{align}
where ${\cal V}_{\mathfrak B'\mathfrak B}$ and ${\cal A}_{\mathfrak B'\mathfrak B}$ are constants
whose values for the aforementioned $\mathfrak B'\mathfrak B$ pairs are collected in
table \ref{VA}, the $u$s are Dirac spinors, \,$\texttt Q=p_{\mathfrak B}^{}-p_{\mathfrak B'}^{}$,\,
with $p_X^{}$ denoting the momentum of $X$,
\begin{align}
{\cal S}_{\mathfrak B'\mathfrak B}^{} & \,=\, \frac{m_{\mathfrak B}^{}-m_{\mathfrak B'}^{}}
{m_s^{}-\hat m}\, {\cal V}_{\mathfrak B'\mathfrak B}^{} \,, &
{\cal P}_{\mathfrak B'\mathfrak B}^{} & \,=\, {\cal A}_{\mathfrak B'\mathfrak B\,}^{} B_0^{}~
\frac{m_{\mathfrak B'}^{}+m_{\mathfrak B}^{}}{m_K^2-\texttt Q^2} \,,
\end{align}
and the other quantities are defined in
appendix \ref{quark-hadron}.
For \,$\Omega^-\to\Xi^-\texttt{\textsl f}\bar{\texttt{\textsl f}}$\, we have
\begin{align} \label{<XO>}
\langle\Xi^-|\overline{d}\gamma^\eta s|\Omega^-\rangle & \,=\, 0 \,, &
\langle\Xi^-\big|\overline{d}\gamma^\eta\gamma_5^{}s|\Omega^-\rangle & \,=\, {\cal C}\,
\bar u_{\Xi}^{} \Bigg( u_{\Omega}^\eta + \frac{\tilde{\textsc q}{}^\eta\,
\tilde{\textsc q}{}_\kappa^{}}{m_K^2-\tilde{\textsc q}{}^2}\, u_\Omega^\kappa \Bigg) , &
\nonumber \\
\langle\Xi^-|\overline{d} s|\Omega^-\rangle & \,=\, 0 \,, &
\langle\Xi^-|\overline{d}\gamma_5^{}s|\Omega^-\rangle & \,=\, \frac{B_0^{}\,{\cal C}\,
\tilde{\textsc q}_\kappa^{}}{\tilde{\textsc q}{}^2-m_K^2}\,\bar u_{\Xi}^{}u_\Omega^\kappa \,,
\end{align}
where \,$\tilde{\textsc q}=p_{\Omega^-}^{}-p_{\Xi^-}^{}$\, and
$u_\Omega^\eta$ is a Rarita-Schwinger spinor.
Accordingly, in our approximation the amplitude for
\,$\Omega^-\to\Xi^-\texttt{\textsl f}\bar{\texttt{\textsl f}}$,\, to be given below, does not
contain the couplings $\texttt C_{\texttt{\textsl f}}^{\texttt V,\texttt A,\texttt S,\texttt P}$.
It is worth noting that the preceding baryonic matrix elements, and their mesonic counterparts
to be discussed in section \ref{kaons}, fulfill the relations
\,$\langle Y|\overline{d}\gamma_\eta^{}s|X\rangle\big(p_X^\eta-p_Y^\eta\big)
= (m_s^{}-\hat m)\langle Y|\overline{d}s|X\rangle$
and
\,$\langle Y|\overline{d}\gamma_\eta^{}\gamma_5^{}s|X\rangle\big(p_Y^\eta-p_X^\eta\big)
= (m_s^{}+\hat m)\langle Y|\overline{d}\gamma_5^{}s|X\rangle$\,
based on the free Dirac equation.

\begin{table}[t]
\begin{tabular}{|c||c|c|c|c|c|c|} \hline
$\mathfrak B'\mathfrak B$ & $n\Lambda$ & $p\Sigma^+$ & $\Lambda\Xi^0$ & $\Sigma^0\Xi^0$ &
$\Sigma^-\Xi^-\vphantom{\int_o^|}$ \\ \hline\hline
~${\cal V}_{\mathfrak B'\mathfrak B}^{}$~ & $-\sqrt{\tfrac{3}{2}}$ & $-1$ &
$\sqrt{\tfrac{3}{2}}^{\vphantom{|}}$ & $\tfrac{-1}{\sqrt2}_{\vphantom{\int}^{}}$ & $1$ \\ \hline
${\cal A}_{\mathfrak B'\mathfrak B}^{}$ &
~$\displaystyle \tfrac{-1}{\sqrt6}_{\vphantom{\int}}^{\vphantom{\int}}(D+3F)$~ & ~$D-F$~ &
~$\displaystyle \tfrac{-1}{\sqrt6}(D-3F)$~ & ~$\displaystyle \tfrac{-1}{\sqrt2}(D+F)$~ &
~$D+F\vphantom{\int_{\int_|}^{\int}}$~ \\ \hline
\end{tabular}
\caption{Values of ${\cal V}_{\mathfrak B'\mathfrak B}$ and ${\cal A}_{\mathfrak B'\mathfrak B}$
in eq.\,(\ref{<B'B>}) for
\,$\mathfrak{BB}'=\Lambda n,\Sigma^+p,\Xi^0\Lambda,\Xi^0\Sigma^0,\Xi^-\Sigma^-$.\,
The constants $D$ and $F$ come from the lowest-order chiral Lagrangian.} \label{VA}
\end{table}

With eq.\,(\ref{<B'B>}), we obtain the amplitude for
\,$\mathfrak B\to\mathfrak B'\texttt{\textsl f}\bar{\texttt{\textsl f}}$\, to be
\begin{align}
{\cal M}_{\mathfrak B\to\mathfrak B'\texttt{\textsl f}\bar{\texttt{\textsl f}}}^{} \,= &~\,
\bar u_{\mathfrak B'}^{} \gamma^\eta u_{\mathfrak B}^{}\, \bar u_{\texttt{\textsl f}}^{}
\gamma_\eta^{} \big( V_{\mathfrak{BB}{}'\texttt{\textsl f}}^{} + \gamma_5^{}
A_{\mathfrak{BB}{}'\texttt{\textsl f}}^{} \big) v_{\bar{\texttt{\textsl f}}}^{}
+ \bar u_{\mathfrak B'}^{}\gamma^\eta\gamma_5^{}u_{\mathfrak B}^{}\, \bar u_{\texttt{\textsl f}}^{}
\gamma_\eta^{} \big( \tilde{\textsc v}_{\mathfrak{BB}{}'\texttt{\textsl f}}^{}
+ \gamma_5^{} \tilde{\textsc a}_{\mathfrak{BB}{}'\texttt{\textsl f}}^{} \big)
v_{\bar{\texttt{\textsl f}}}^{}
\nonumber \\ & \,+\,\!
\bar u_{\mathfrak B'}^{}u_{\mathfrak B}^{}\, \bar u_{\texttt{\textsl f}}^{} \big(
S_{\mathfrak{BB}{}'\texttt{\textsl f}}^{} + \gamma_5^{} P_{\mathfrak{BB}{}'\texttt{\textsl f}}^{}
\big) v_{\bar{\texttt{\textsl f}}}^{} + \bar u_{\mathfrak B'}^{} \gamma_5^{}u_{\mathfrak B}^{}\,
\bar u_{\texttt{\textsl f}}^{} \big( \tilde{\textsc s}_{\mathfrak{BB}{}'\texttt{\textsl f}}^{}
+ \gamma_5^{} \tilde{\textsc p}_{\mathfrak{BB}{}'\texttt{\textsl f}}^{} \big)
v_{\bar{\texttt{\textsl f}}}^{} \,,
\end{align}
where the $v$s are Dirac spinors for $\bar{\texttt{\textsl f}}$,
\begin{align} \label{vasp}
V_{\mathfrak{BB}{}'\texttt{\textsl f}}^{} & = {\cal V}_{\mathfrak B'\mathfrak B}^{}
{\texttt C}_{\texttt{\textsl f}}^{\texttt V} , &
A_{\mathfrak{BB}{}'\texttt{\textsl f}}^{} & = {\cal V}_{\mathfrak B'\mathfrak B}^{}
{\texttt C}_{\texttt{\textsl f}}^{\texttt A} , &
\tilde{\textsc v}_{\mathfrak{BB}{}'\texttt{\textsl f}}^{} & =
{\cal A}_{\mathfrak B'\mathfrak B\,}^{} \tilde{\textsf c}_{\texttt{\textsl f}}^{\textsc v} , &
\tilde{\textsc a}_{\mathfrak{BB}{}'\texttt{\textsl f}}^{} & =
{\cal A}_{\mathfrak B'\mathfrak B\,}^{} \tilde{\textsf c}_{\texttt{\textsl f}}^{\textsc a} ,
\nonumber \\
S_{\mathfrak{BB}{}'\texttt{\textsl f}}^{} & = {\cal S}_{\mathfrak B'\mathfrak B}^{}
{\texttt C}_{\texttt{\textsl f}}^{\texttt S} , &
P_{\mathfrak{BB}{}'\texttt{\textsl f}}^{} & = {\cal S}_{\mathfrak B'\mathfrak B}^{}
{\texttt C}_{\texttt{\textsl f}}^{\texttt P} , &
\tilde{\textsc s}_{\mathfrak{BB}{}'\texttt{\textsl f}}^{} & =
{\cal P}_{\mathfrak B'\mathfrak B\,}^{} \tilde{\textsf c}_{\texttt{\textsl f}}^{\textsc s} , &
\tilde{\textsc p}_{\mathfrak{BB}{}'\texttt{\textsl f}}^{} & =
{\cal P}_{\mathfrak B'\mathfrak B\,}^{} \bigg( \tilde{\textsf c}_{\texttt{\textsl f}}^{\textsc p}
- \frac{2m_{\texttt{\textsl f}}^{}}{B_0}\,\tilde{\textsf c}_{\texttt{\textsl f}}^{\textsc a}\bigg) .
\end{align}
This leads to the differential decay rate
\begin{align} \label{G'B2B'em}
\frac{d\Gamma_{\mathfrak B\to\mathfrak B'\texttt{\textsl f}\bar{\texttt{\textsl f}}}}{d\hat s} \,=\,
\frac{\beta\, \lambda_{\mathfrak{BB}'}^{1/2}}{128\pi^3 m_{\mathfrak B}^3} &
\begin{array}[t]{l} \!
\Big[ \big( \beta^2 \tilde\sigma_+^{}\hat s + 3 \tilde\sigma_-^{} \hat s + {\mathbb F} \big)
\big|V_{\mathfrak{BB}{}'\texttt{\textsl f}}^{}\big|^2 + \big( \tilde\sigma_+^{} \hat s
+ 3\beta^2 \tilde\sigma_-^{} \hat s + {\mathbb F} \big)
\big|A_{\mathfrak{BB}{}'\texttt{\textsl f}}^{}\big|^2
\vspace{4pt} \\ +\,
\big( \beta^2 \tilde\sigma_-^{} \hat s + 3 \tilde\sigma_+^{} \hat s + {\mathbb F} \big)
\big|\tilde{\textsc v}_{\mathfrak{BB}{}'\texttt{\textsl f}}^{}\big|^2
+ \big( \tilde\sigma_-^{} \hat s + 3 \beta^2 \tilde\sigma_+^{} \hat s + {\mathbb F} \big)
\big|\tilde{\textsc a}_{\mathfrak{BB}{}'\texttt{\textsl f}}^{}\big|^2
\vspace{4pt} \\ +~
\tilde\sigma_+^{} \Big( \beta^2 \big|S_{\mathfrak{BB}{}'\texttt{\textsl f}}^{}\big|^2 \!
+ \big|P_{\mathfrak{BB}{}'\texttt{\textsl f}}^{}\big|^2 \Big) \hat s
+ \tilde\sigma_-^{} \Big( \beta^2
\big|\tilde{\textsc s}_{\mathfrak{BB}{}'\texttt{\textsl f}}^{}\big|^2 \!
+ \big|\tilde{\textsc p}_{\mathfrak{BB}{}'\texttt{\textsl f}}^{}\big|^2 \Big) \hat s \end{array}
\nonumber \\ & +\,
4 m_{\texttt{\textsl f}}^{}\, {\rm Re} \Bigl( \tilde\sigma_+^{}\, {\texttt M}_-^{}\,
A_{\mathfrak{BB}{}'\texttt{\textsl f}\,}^* P_{\mathfrak{BB}{}'\texttt{\textsl f}}^{}
- \tilde\sigma_-^{}\, {\texttt M}_+^{}\, \tilde{\textsc a}{}_{\mathfrak{BB}{}'\texttt{\textsl f}\,}^*
\tilde{\textsc p}_{\mathfrak{BB}{}'\texttt{\textsl f}}^{} \Bigr) \Big] \,,
\end{align}
where
\begin{align}
\beta & \,=\, \sqrt{1-\frac{4 m_{\texttt{\textsl f}}^2}{\hat s}} \,, &
\hat s & \,=\, \big(p_{\texttt{\textsl f}}^{}+p_{\bar{\texttt{\textsl f}}}^{}\big)^2 \,, &
\lambda_{XY}^{} & \,=\, m_X^4-2\big(m_Y^2+\hat s\big)m_X^2
+ \big(m_Y^2-\hat s\big)\raisebox{1pt}{$^2$} \,,
\nonumber \\
\tilde\sigma_\pm^{} & \,=\, {\texttt M}_\pm^2 - \hat s \,, &
{\texttt M}_\pm^{} & \,=\, m_{\mathfrak B}^{}\pm m_{\mathfrak B'}^{} \,, &
{\mathbb F} & \,=\, \frac{3-\beta^2}{3}\lambda_{\mathfrak{BB}{}'}^{}
+ 2\beta^2 \big(\hat s-m_{\mathfrak B}^2-m_{\mathfrak B'}^2\big)\hat s \,.
\end{align}
The rate results from integrating the differential rate over
\,$4m_{\texttt{\textsl f}}^2\le\hat s\le(m_{\mathfrak B}-m_{\mathfrak B'})^2$.\,
In eq.\,(\ref{G'B2B'em}) we observe that the
$\texttt C_{\texttt{\textsl f}}^{\texttt V,\texttt A,\texttt S,\texttt P}$
terms do not interfere with the
$\tilde{\textsf c}_{\texttt{\textsl f}}^{\textsc v,\textsc a,\textsc s,\textsc p}$
ones, which is also the case in the kaon decays to be examined later on.

For the $\Omega^-$ decay we find
\begin{align}
{\cal M}_{\Omega^-\to\Xi^-\texttt{\textsl f}\bar{\texttt{\textsl f}}}^{} \,=\, {\cal C}_{\,}
\Bigg( g_{\kappa\varsigma}^{} + \frac{\tilde{\textsc q}_\kappa^{}\tilde{\textsc q}_\varsigma^{}}
{m_{K^0}^2-\hat s} \Bigg) \bar u_\Xi^{} u_\Omega^\kappa\, \bar u_{\texttt{\textsl f}}^{}
\Bigg[ \gamma^\varsigma \big( \tilde{\textsf c}_{\texttt{\textsl f}}^{\textsc v}
+ \gamma_5^{} \tilde{\textsf c}_{\texttt{\textsl f}}^{\textsc a} \big) - \frac{B_{0\,}^{}
\tilde{\textsc q}{}^\varsigma}{m_{K^0}^2} \big(
\tilde{\textsf c}_{\texttt{\textsl f}}^{\textsc s} + \gamma_5^{}
\tilde{\textsf c}_{\texttt{\textsl f}}^{\textsc p} \big) \Bigg]v_{\bar{\texttt{\textsl f}}}^{} \,, &
\end{align}
where \,$\tilde{\textsc q}=p_{\Omega^-}^{}-p_{\Xi^-}^{}$.\,
The resulting differential rate is
\begin{align} \label{G'O2Xem}
\frac{d\Gamma_{\Omega^-\to\Xi^-\texttt{\textsl f}\bar{\texttt{\textsl f}}}}{d\hat s} \,=\, \frac{\beta\,
\lambda_{\Omega^-\Xi^-\,}^{1/2} {\cal C}^2\hat s}{1536\pi^{3\,} m_{\Omega^-}^5} &
\Big\{ \big(3-\beta^2\big) {\cal G}\, \big|\tilde{\textsf c}_{\texttt{\textsl f}}^{\textsc v}\big|^2
+ \Big[ 2\beta^2 {\cal  G} + \big(1-\beta^2\big) \tilde K\, m_{K^0}^4 \Big]
\big|\tilde{\textsf c}_{\texttt{\textsl f}}^{\textsc a}\big|^2
\nonumber \\ & \,+\;\!
B_0^2 \Big( \beta^{2\,} \big|\tilde{\textsf c}_{\texttt{\textsl f}}^{\textsc s}\big|^2
+ \big|\tilde{\textsf c}_{\texttt{\textsl f}}^{\textsc p}\big|^2 \Big) \tilde K \hat s - 4 B_0^{} \tilde K
m_{\texttt{\textsl f}\,}^{} m_{K^0}^2\, {\rm Re} \big( \tilde{\textsf c}_{\texttt{\textsl f}}^{\textsc a*}
\tilde{\textsf c}_{\texttt{\textsl f}}^{\textsc p} \big) \Big\} \,,
\end{align}
where
\begin{align}
{\cal  G} & \,=\, \bigg( \frac{\lambda_{\Omega^-\Xi^-}}{3} + 4 m_{\Omega^-}^2\hat s
\bigg) \frac{(m_{\Omega^-}+m_{\Xi^-})^2-\hat s}{\hat s} \,, &
\tilde K & \,=\, \lambda_{\Omega^-\Xi^-}\, \frac{(m_{\Omega^-}+m_{\Xi^-})^2-\hat s}
{\big(m_{K^0}^2-\hat s\big)\raisebox{1pt}{$^2$} \hat s} \,.
\end{align}

In eq.\,(\ref{<B'B>}) there are form-factor effects not yet taken into account.
To incorporate them, in numerical work we modify ${\cal V}_{\mathfrak B'\mathfrak B}$ and
${\cal A}_{\mathfrak B'\mathfrak B}$ to
\,$\big(1+2\texttt Q^2/M_V^2\big){\cal V}_{\mathfrak B'\mathfrak B}$\, and
\,$\big(1+2\texttt Q^2/M_A^2\big){\cal A}_{\mathfrak B'\mathfrak B}$,\, respectively, with
\,$M_V=0.97$ GeV and \,$M_A=1.25$ GeV,\, following the commonly used parametrization in
experimental analyses of hyperon semileptonic decays
\cite{Bourquin:1981ba,Hsueh:1988ar,Dworkin:1990dd,Batley:2006fc} and assuming isospin symmetry.
Analogously, since $\tilde{\textsc q}{}^2$ in
\,$\Omega^-\to\Xi^-\texttt{\textsl f}\bar{\texttt{\textsl f}}$\, has a significantly wider
range than $\texttt Q^2$ in
$\mathfrak B\to\mathfrak B'\texttt{\textsl f}\bar{\texttt{\textsl f}}$, in the $\Omega^-$ decay
rate we implement the change
\,${\cal C}\to{\cal C}/ \big(1-\tilde{\textsc q}{}^2/M_A^2\big)\raisebox{1pt}{$^2$}$.\,
These modifications turn out to translate into increases of the rates by up to
{\footnotesize\,$\sim$\,}16 percent.

\section{Kaon decays\label{kaons}}

For \,$K_{L,S}\to\texttt{\textsl f}\bar{\texttt{\textsl f}}$\, the relevant hadronic
matrix elements are
\begin{align} \label{<0K>}
\langle0|\overline{d}\gamma^\eta\gamma_5^{}s|\,\overline{\!K}{}^0\rangle & =
\langle0|\overline{s}\gamma^\eta\gamma_5^{}d|K^0\rangle =\, -i f_K^{} p_K^\eta \,, &
\langle0|\overline{d}\gamma_5^{}s|\,\overline{\!K}{}^0\rangle & =
\langle0|\overline{s}\gamma_5^{}d|K^0\rangle =\, i B_0^{}f_K^{} \,,
\end{align}
with \,$f_K^{}=155.6(4)$\,MeV \cite{Tanabashi:2018oca} being the kaon decay constant, while for
\,$K\to\pi\texttt{\textsl f}\bar{\texttt{\textsl f}}$\,
\begin{align}
\langle\pi^-|\bar d\gamma^\eta s|K^-\rangle & \,=\, -\langle\pi^+|\bar s\gamma^\eta d|K^+\rangle
\,=\, \big(p_K^\eta+p_\pi^\eta\big) f_+^{} \,+\, \big(f_0^{}-f_+^{}\big) q_{K\pi}^\eta\,
\frac{m_K^2-m_\pi^2}{q_{K\pi}^2} \,, &
\nonumber \\ \label{<piK>}
\langle\pi^-|\bar d s|K^-\rangle & \,=\, +\langle\pi^+|\bar s d|K^+\rangle \,=\, B_0^{} f_0^{} \,,
~~~ ~~~~ ~~~ q_{K\pi}^{} \,=\, p_K^{}-p_\pi^{} \,,
\end{align}
where $f_+^{}$ and $f_0^{}$ represent form factors which are functions of $q_{K\pi}^2$.
In addition, assuming isospin symmetry, we have
\,$\big\langle\pi^0\big|\bar d(\gamma^\eta,1)s\big|\,\overline{\!K}{}^0\big\rangle =
\big\langle\pi^0\big|\bar s(-\gamma^\eta,1)d\big|K^0\big\rangle =
-\big\langle\pi^-\big|\bar d(\gamma^\eta,1)s\big|K^-\big\rangle/\sqrt2$\,
and also
\,$\big\langle\pi^-\big|\bar d\gamma^\eta s\big|K^-\big\rangle=
\big\langle\pi^+\big|\bar u\gamma^\eta s\big|\,\overline{\!K}{}^0\big\rangle$.\,
We can then adopt
$f_{+,0}^{}=\textsf f_+^{}(0)\big(1+\lambda_{+,0}^{}\,q_{K\pi}^2/m_{\pi^+}^2\big)$\,
with \,$\lambda_+^{}=0.0271(10)$\, and \,$\lambda_0^{}=0.0142(23)$\, from
\,$K_L\to\pi^+\mu^-\nu$ measurements \cite{Tanabashi:2018oca} as well as \,$\textsf f_+^{}(0)=0.9681(23)$\,
from lattice computations \cite{Charles:2015gya}.\footnote{Online updates available at
http://ckmfitter.in2p3.fr.}
For \,$K^-\to\pi^0\pi^-\texttt{\textsl f}\bar{\texttt{\textsl f}}$\, and
\,$K_L\to\pi^0\pi^0\texttt{\textsl f}\bar{\texttt{\textsl f}}$,\, from the results in
appendix \ref{quark-hadron} we obtain
\begin{align} \label{<K->pp>}
\big\langle\pi^0(p_0^{})\,\pi^-(p_-^{})\big|\bar d\big(\gamma^\eta,1\big)\gamma_5^{}s
\big|K^-\big\rangle
& \,=\, \frac{i\sqrt2}{f_K^{}} \bigg[ \big( p_0^\eta - p_-^\eta, 0 \big)
+ \frac{(p_0^{}-p_-^{})\cdot\tilde{\textsl{\texttt q}}}{m_K^2-\tilde{\textsl{\texttt q}}{}^2}
\bigl( \tilde{\textsl{\texttt q}}{}^\eta, -B_0^{} \bigr) \bigg] \,, &
\nonumber \\
\big\langle\pi^0(p_1^{})\,\pi^0(p_2^{})\big|\bar d\big(\gamma^\eta,1\big)\gamma_5^{}s
\big|\,\overline{\!K}{}^0\big\rangle
& \,=\, \frac{i}{f_K^{}} \bigg[ \big( p_1^\eta + p_2^\eta, 0 \big)
+ \frac{(p_1^{}+p_2^{})\cdot\tilde{\textsl{\texttt q}}}{m_K^2-\tilde{\textsl{\texttt q}}{}^2}
\bigl( \tilde{\textsl{\texttt q}}{}^\eta, -B_0^{} \bigr) \bigg] \,, &
\end{align}
where \,$\tilde{\textsl{\texttt q}}=p_{K^-}^{}-p_0^{}-p_-^{}=p_{\bar K^0}^{}-p_1^{}-p_2^{}$.\,
In the $K^-$ case, there is additionally a small contribution involving one of the parity-even quark transitions, \,$\langle\pi^0\pi^-|\bar d\gamma^\eta s|K^-\rangle\neq0$,\, which arises from the anomaly Lagrangian \cite{Kamenik:2011vy}, at next-to-leading order in the chiral expansion, and which we have therefore neglected.
Since the existing empirical limits on \,$K\to\pi\pi'\texttt{\textsl f}\bar{\texttt{\textsl f}}$\,
are not very stringent, we also ignore form-factor effects in calculating their rates.

It follows that the amplitudes for \,$K_L\to\texttt{\textsl f}\bar{\texttt{\textsl f}}$\, and
\,$K_S\to\texttt{\textsl f}\bar{\texttt{\textsl f}}$\, induced by ${\cal L}_{\texttt{\textsl f}}$
are
\begin{align} \label{MK2ff}
{\cal M}_{K_{L,S\,}^{}\to\texttt{\textsl f}\bar{\texttt{\textsl f}}}^{} & \,=\,
i\, \bar u_{\texttt{\textsl f}}^{} \Big( S_{K_{L,S\,}\texttt{\textsl f}}^{}
+ \gamma_5^{} P_{K_{L,S\,}\texttt{\textsl f}}^{} \Big) v_{\bar{\texttt{\textsl f}}}^{} \,, &
\end{align}
leading to the decay rates
\begin{align} \label{GK2ff}
\Gamma_{K_{L,S}\to\texttt{\textsl f}\bar{\texttt{\textsl f}}}^{} & \,=\,
\frac{m_{K^0}^{}}{8\pi} \Big( \tilde\beta^3
\big|S_{K_{L,S\,}\texttt{\textsl f}} \big| \raisebox{2pt}{$^2$} + \tilde\beta\,
\big| P_{K_{L,S\,}\texttt{\textsl f}}^{} \big| \raisebox{2pt}{$^2$} \Big) \,, &
\end{align}
where
\,$\tilde\beta=\big(1-4m_{\texttt{\textsl f}}^2/m_{K^0}^2\big)\raisebox{1pt}{$^{1/2}$}$,\,
\begin{align} \label{SK2ff}
S_{K\!_L^{}\texttt{\textsl f}}^{} & \,=\,
i\sqrt2\, B_{0\,}^{} f_K^{}\, {\rm Im}\, \tilde{\textsf c}_{\texttt{\textsl f}}^{\textsc s} \,, &
P_{K\!_L^{}\texttt{\textsl f}}^{} & \,=\, -\sqrt2\, f_K^{}\,
{\rm Re} \big( 2 m_{\texttt{\textsl f}}^{}\, \tilde{\textsf c}_{\texttt{\textsl f}}^{\textsc a}
- B_0^{}\, \tilde{\textsf c}_{\texttt{\textsl f}}^{\textsc p} \bigr) \,,
\nonumber \\
S_{K\!_S^{}\texttt{\textsl f}}^{} & \,=\,
-\sqrt2\, B_{0\,}^{} f_K^{}\, {\rm Re}\, \tilde{\textsf c}_{\texttt{\textsl f}}^{\textsc s} \,, &
P_{K\!_S^{}\texttt{\textsl f}}^{} & \,=\, i\sqrt2\, f_K^{}\,
{\rm Im} \big( 2 m_{\texttt{\textsl f}}^{}\, \tilde{\textsf c}_{\texttt{\textsl f}}^{\textsc a}
- B_0^{}\, \tilde{\textsf c}_{\texttt{\textsl f}}^{\textsc p} \big) \,.
\end{align}
Thus, \,$K_{L,S}\to\texttt{\textsl f}\bar{\texttt{\textsl f}}$\, are not sensitive to
$\texttt C_{\texttt{\textsl f}}^{\texttt V,\texttt A,\texttt S,\texttt P}$ and
$\tilde{\textsf c}_{\texttt{\textsl f}}^{\textsc v}$.

The amplitude for \,$K\to\pi\texttt{\textsl f}\bar{\texttt{\textsl f}}$\, has the form
\begin{equation} \label{MK2pff}
{\cal M}_{K\to\pi\texttt{\textsl f}\bar{\texttt{\textsl f}}} \,=\,
\bar u_{\texttt{\textsl f}}^{} \big( S_{K\pi\texttt{\textsl f}}^{}
+ P_{K\pi\texttt{\textsl f}\,}^{} \gamma_5^{} \big) v_{\bar{\texttt{\textsl f}}}^{} \,.
\end{equation}
We put the resulting differential rates of
\,$K^-\to\pi^-\texttt{\textsl f}\bar{\texttt{\textsl f}}$\, and
\,$K_{L,S}\to\pi^0\texttt{\textsl f}\bar{\texttt{\textsl f}}$\, in
appendix \ref{Kformulas}, which also shows that these modes, in contrast to
\,$K_{L,S}\to\texttt{\textsl f}\bar{\texttt{\textsl f}}$,\, can probe
$\texttt C_{\texttt{\textsl f}}^{\texttt V,\texttt A,\texttt S,\texttt P}$, but not
$\tilde{\textsf c}_{\texttt{\textsl f}}^{\textsc v,\textsc a,\textsc s,\textsc p}$.

For \,$K^-\to\pi^0\pi^-\texttt{\textsl f}\bar{\texttt{\textsl f}}$\, and
\,$K_L\to\pi^0\pi^0\texttt{\textsl f}\bar{\texttt{\textsl f}}$,\, we get
\begin{align}
{\cal M}_{K^-\to\pi^0\pi^-\texttt{\textsl f}\bar{\texttt{\textsl f}}}^{} & =
\frac{i\sqrt2}{f_K^{}}\, \bar u_{\texttt{\textsl f}}^{} \bigg\{ \!
\big[ 2\gamma_5^{} \tilde{\textsf c}_{\texttt{\textsl f}}^{\textsc a} m_{\texttt{\textsl f}}^{}
- B_0^{} \big( \tilde{\textsf c}_{\texttt{\textsl f}}^{\textsc s} + \gamma_5^{}
\tilde{\textsf c}_{\texttt{\textsl f}}^{\textsc p} \big) \big]
\frac{p_K^{}\!\cdot\!(p_0^{}-p_-^{})}{m_K^2-\hat s}
+ \big(\slashed p{}_0^{}-\slashed p{}_-^{}\big) \big( \tilde{\textsf c}_{\texttt{\textsl f}}^{\textsc v}
+ \gamma_5^{} \tilde{\textsf c}_{\texttt{\textsl f}}^{\textsc a} \big) \! \bigg\}
v_{\bar{\texttt{\textsl f}}\,}^{} ,
\nonumber \\
{\cal M}_{K_L\to\pi^0\pi^0\texttt{\textsl f}\bar{\texttt{\textsl f}}}^{} & =
\frac{i\sqrt2}{f_K^{}}\, \bar u_{\texttt{\textsl f}}^{} \bigg\{  \begin{array}[t]{l} \!\! \big[ 2
\gamma_5^{}\,{\rm Re}\,\tilde{\textsf c}_{\texttt{\textsl f}}^{\textsc a\,}m_{\texttt{\textsl f}}^{}
- B_0^{} \big( i\, {\rm Im}\,\tilde{\textsf c}_{\texttt{\textsl f}}^{\textsc s} + \gamma_5^{}\,
{\rm Re}\, \tilde{\textsf c}_{\texttt{\textsl f}}^{\textsc p} \big) \big]   \displaystyle
\frac{p_K^{}\!\cdot\!\big(p_{\textsl{\texttt f}}^{}+p_{\bar{\textsl{\texttt f}}}^{}\big)
- \hat s}{m_K^2-\hat s}
\\ \!\!+\;   \displaystyle
\big(\slashed p{}_1^{} + \slashed p{}_2^{}\big) \big(  \displaystyle
{\rm Re}\, \tilde{\textsf c}_{\texttt{\textsl f}}^{\textsc v}
+ \gamma_5^{}\, {\rm Re}\, \tilde{\textsf c}_{\texttt{\textsl f}}^{\textsc a} \big)
\bigg\} v_{\bar{\texttt{\textsl f}}}^{} \,. \end{array}
\end{align}
Their differential rates are also relegated to appendix \ref{Kformulas}.

We notice from eqs.\,\,(\ref{vasp}) and (\ref{G'B2B'em}) that, unlike these kaon modes,
\,$\mathfrak B\to\mathfrak B'\texttt{\textsl f}\bar{\texttt{\textsl f}}$\,
are sensitive to both $\texttt C_{\texttt{\textsl f}}^{\texttt V,\texttt A,\texttt S,\texttt P}$
and $\tilde{\textsf c}_{\texttt{\textsl f}}^{\textsc v,\textsc a,\textsc s,\textsc p}$.
It is therefore advantageous to measure \,$\mathfrak B\to\mathfrak B'\slashed E$,\, as the acquired
data could supply information on \,$s\to d\slashed E$\, which is complementary to that from
the kaon sector.

\section{Numerical results\label{numeric}}

\subsection{SM predictions and empirical information\label{smx}}

Within the SM, our hyperon decays of interest are induced by effective short-distance
\,$sd\nu_l\bar\nu_l$\, interactions, with \,$l=e,\mu,\tau$,\, described by \cite{Buchalla:1995vs}
\begin{align} \label{Lsm}
{\cal L}_{sd\nu\nu}^{\textsc{sm}} & \,=\, \frac{-\alpha_{\rm e}^{}G_{\rm F}^{}}{\sqrt8\,\pi
s_{\textsc w}^2}\, \raisebox{2pt}{\footnotesize$\displaystyle\sum_{l=e,\mu,\tau}$}
\big( V_{td}^*V_{ts}^{} X_t^{}+V_{cd}^*V_{cs}^{} X_c^l\big)\, \overline{d}\gamma^\eta
(1-\gamma_5^{})s~ \overline{\nu_l^{}}\gamma_\eta^{}(1-\gamma_5^{})\nu_l^{}
\;+\; {\rm H.c.} \,, &
\end{align}
where \,$\alpha_{\rm e}^{}\simeq1/128$\, and $G_{\rm F}$ are the usual fine-structure and Fermi
constants, \,$s_{\textsc w}^2\equiv\sin^2\!\theta_{\textsc w}=0.231$ with $\theta_{\textsc w}$
being the Weinberg angle, $V_{qq'}$ are Cabibbo-Kobayashi-Maskawa (CKM) matrix elements,
\,$X_t^{}=1.481(9)$\, comes from $t$-quark loops \cite{Buras:2015qea}, and
\,$X_c^e=X_c^\mu\simeq1.0\times10^{-3}$ and \,$X_c^\tau\simeq7\times10^{-4}$ are $c$-quark
contributions \cite{Buchalla:1995vs}.
To evaluate
\,${\cal B}(\mathfrak B\to\mathfrak B'\nu\bar\nu)_{\textsc{sm}}^{}=\raisebox{1pt}{\small$\sum$}_l^{}
{\cal B}(\mathfrak B\to\mathfrak B'\nu_l\bar\nu_l)_{\textsc{sm}}^{}$
we can apply eq.\,(\ref{G'B2B'em}) with
\,$\texttt C_{\nu_l}^{\texttt V}=-\texttt C_{\nu_l}^{\texttt A}
= -\tilde{\textsf c}_{\nu_l}^{\textsc v}
= \tilde{\textsf c}_{\nu_l}^{\textsc a}=\alpha_{\rm e}^{}G_{\rm F}^{}
\big(\lambda_t^{}X_t^{}+ \lambda_c^{}X_c^l\big)/\big(\sqrt8\,\pi s_{\textsc w}^2\big)$
and \,$\texttt C_{\nu_l}^{\texttt S,\texttt P}=\tilde{\textsf c}_{\nu_l}^{\textsc s,\textsc p}=0$.
Similarly, we can determine \,${\cal B}(\Omega^-\to\Xi^-\nu\bar\nu)_{\textsc{sm}}^{}$\,
using eq.\,(\ref{G'O2Xem}).\footnote{These decays also receive long-distance contributions
mediated by the $Z$ boson, such as $\mathfrak B\to\mathfrak B'Z^*\to\mathfrak B'\nu\bar\nu$,
but their size is estimated to be small compared to the short-distance SM contribution.\medskip}
Thus, with the central values of the input parameters, we arrive at the entries in the second
row of table \ref{smB2B'nunu}.\footnote{Our numbers are roughly comparable to the ones given
in \cite{Hu:2018luj}, but therein the flavor-SU(3) properties of the baryon interactions were not
taken into account and different momentum-dependences of the baryonic matrix elements were used.}
The CKM factors and $X_{t,c}$ contribute an uncertainty of almost 10\% to the SM predictions,
the estimation of the baryonic matrix elements has an uncertainty of {\small$\sim$\,}20\%,
and so the total uncertainty of the branching-fraction predictions is about 50\%.

At present there are no data available on these hyperon transitions, but this situation may
change in the near future if BESIII performs a quest for them.
In the last row of table \ref{smB2B'nunu} we quote its estimated sensitivities
[90\% confidence level (CL)] for their branching fractions \cite{Li:2016tlt}.
Clearly, it is unlikely that the SM predictions will be tested anytime soon.
Nevertheless, as we demonstrate in the next section, it is possible for NP to amplify
the branching fractions to levels potentially reachable by BESIII.

\begin{table}[b] \bigskip
\begin{tabular}{|c||c|c|c|c|c|c|} \hline
Decay mode & \,$\Lambda\to n\nu\bar\nu$\, & \,$\Sigma^+\to p\nu\bar\nu$\, &
\,$\Xi^0\to\Lambda\nu\bar\nu$\, & \,$\Xi^0\to\Sigma^0\nu\bar\nu$\, &
\,$\Xi^-\to\Sigma^-\nu\bar\nu$\, & \,$\Omega^-\to\Xi^-\nu\bar\nu\vphantom{\int^|}$\,
\\ \hline \hline
\footnotesize $\begin{array}{c}\rm SM ~branching\vspace{-1ex}\\ \rm fraction \end{array}$ &
\,$7.1\times10^{-13}$\, & \,$4.3\times10^{-13}$\, & \,$6.3\times10^{-13}$\, &
$1.0\times10^{-13}$ & $1.3\times10^{-13}$ & $4.9\times10^{-12}$
\\ \hline
\footnotesize $\begin{array}{c}\rm Expected~BESIII\vspace{-1ex}\\
\rm sensitivity~\mbox{\cite{Li:2016tlt}}\end{array}$ &
$3\times10^{-7}$ & $4\times10^{-7}$ & $8\times10^{-7}$\, & $9\times10^{-7}$ & {\bf---} &
$2.6\times10^{-5}$ \\ \hline
\end{tabular}
\caption{The branching fractions of \,$|\Delta S|=1$\, hyperon decays with missing energy in
the SM and the corresponding expected sensitivities of BESIII \cite{Li:2016tlt}.\label{smB2B'nunu}}
\end{table}

Turning to the kaon sector, we see that eqs.\,\,(\ref{GK2ff}) and (\ref{SK2ff}) imply
\,$\Gamma_{K_{L,S}\to\nu\bar\nu}^{\textsc{sm}}=0$\, due to the neutrinos' masslessness in the SM.
If it is supplemented with nonzero neutrino masses, their highest one from the direct limit
\,$m_{\nu_\tau}^{\rm exp}<18.2$\,MeV \cite{Tanabashi:2018oca} translates into the maximal values
\,${\cal B}(K_L\to\nu\bar\nu)_{\textsc{sm}}^{}\simeq1\times10^{-10}$\, and
\,${\cal B}(K_S\to\nu\bar\nu)_{\textsc{sm}}^{}\simeq2\times10^{-14}$.\,
Therefore, observations of \,${\cal B}(K_{L,S}\to\slashed E)\gg10^{-10}$\, would constitute
evidence of NP.
Although to date there are still no measurements on \,$K_{L,S}\to\slashed E$,\, from
the available data \cite{Tanabashi:2018oca} on the visible decay modes of $K_{L,S}$ one can extract
indirect upper bounds on their invisible branching fractions \cite{Gninenko:2014sxa}:
\begin{align} \label{K2inv}
{\mathcal B}(K_L\to\slashed E) & \,<\, 6.3\times10^{-4} \,, &
{\mathcal B}(K_S\to\slashed E) & \,<\, 1.1\times10^{-4} &
\end{align}
both at 95\% CL.
Hence there is still plenty of room for NP to influence these decays, particularly via
the couplings $\tilde{\textsf c}_{\texttt{\textsl f}}^{\textsc a,\textsc s,\textsc p}$,
as eq.\,(\ref{SK2ff}) indicates.

For the \,$K\to\pi\nu\bar\nu$\, modes, the SM predictions are
\,${\mathcal B}(K^+\to\pi^+\nu\nu)_{\textsc{sm}}^{}=\big(8.5_{-1.2}^{+1.0}\big)\times10^{-11}$ and
\,${\mathcal B}(K_L\to\pi^0\nu\bar\nu)_{\textsc{sm}}^{}=\big(3.2_{-0.7}^{+1.1}\big)\times10^{-11}$
\cite{Bobeth:2017ecx}.
These are not very far from their measurements
${\mathcal B}(K^+\to\pi^+\nu\nu)_{\rm exp}^{}=1.7(1.1)\times10^{-10}$ \cite{Tanabashi:2018oca}
and
${\mathcal B}\big(K_L\to\pi^0\nu\bar\nu\big)_{\rm exp}^{}<3.0\times10^{-9}$ at 90\% CL
\cite{Ahn:2018mvc}.
It follows that the effects of NP on these modes, and consequently its contributions to
$\texttt C_{\texttt{\textsl f}}^{\texttt V,\texttt A,\texttt S,\texttt P}$, expectedly
cannot be considerable.

As regards the four-body kaon decays, the SM expectations are
\,${\mathcal B}(K^-\to\pi^0\pi^-\nu\bar\nu)_{\textsc{sm}}^{}\sim10^{-14}$\, and
\,${\mathcal B}(K_L\to\pi^0\pi^0\nu\bar\nu)_{\textsc{sm}}^{}\sim10^{-13}$
\cite{Kamenik:2011vy,Littenberg:1995zy,Chiang:2000bg}.
These are way below the existing empirical bounds
${\mathcal B}(K^-\to\pi^0\pi^-\nu\bar\nu)_{\rm exp}^{}<4.3\times10^{-5}$
\cite{Adler:2000ic}\footnote{The search region was defined by
\,$90{\rm\,MeV}<P_{\pi^-}<188$\,MeV\, and \,$135{\rm\,MeV}<E_{\pi^0}<180$\,MeV
\cite{Adler:2000ic}.} and
${\mathcal B}(K_L\to\pi^0\pi^0\nu\bar\nu)_{\rm exp}^{}<8.1\times10^{-7}$ \cite{E391a:2011aa},
both at 90\%~CL.
We may then impose
\begin{align} \label{K2ppinv}
{\mathcal B}(K^-\to\pi^0\pi^-\slashed E) & \,<\, 4\times10^{-5} \,, &
{\mathcal B}(K_L\to\pi^0\pi^0\slashed E) & \,<\, 8\times10^{-7} \,, &
\end{align}
which imply further restraints on
$\tilde{\textsf c}_{\texttt{\textsl f}}^{\textsc v,\textsc a,\textsc s,\textsc p}$.

\subsection{Beyond SM\label{bsm}}

As mentioned in the preceding subsection, the current \,$K\to\pi\nu\bar\nu$\, data
do not leave ample room for NP to affect
$\texttt C_{\texttt{\textsl f}}^{\texttt V,\texttt A,\texttt S,\texttt P}$ greatly.
More specifically, our numerical scans reveal that the allowed values of these couplings
alone cannot produce
${\cal B}\big(\mathfrak B\to\mathfrak B'\texttt{\textsl f}\bar{\texttt{\textsl f}}\big)$
above $10^{-11}$, and so this scenario would be out of BESIII reach according to
table \ref{smB2B'nunu}.

Therefore, hereafter we concentrate on the possibility that NP can generate sizable effects only
via $\tilde{\textsf c}_{\texttt{\textsl f}}^{\textsc v,\textsc a,\textsc s,\textsc p}$.
For simplicity, we assume that $\texttt{\textsl f}$ is a nonstandard fermion which is
sufficiently light compared to the mass difference between the initial and final baryons,
so that we can approximately set $m_{\texttt{\textsl f}}$ to zero in numerical work.
It follows that only the constraints on \,$K\to\texttt{\textsl f}\bar{\texttt{\textsl f}}$ and
$K\to\pi\pi'\texttt{\textsl f}\bar{\texttt{\textsl f}}$\, need to be addressed when dealing with
the hyperon decays.
Moreover, since the NP contributions do not interfere with \,$s\to d\nu\bar\nu$,\,
the tiny SM contributions to these processes with missing energy can be ignored.

Integrating the differential rates of the baryon decays,
for \,$m_{\texttt{\textsl f}}^{}=0$\, we arrive at the branching fractions
\begin{align} \label{B2B'ff}
{\cal B}\big(\Lambda\to n\texttt{\textsl f}\bar{\texttt{\textsl f}}\big) & \,=
\begin{array}[t]{l} \! \Big[ 4.4\, \Big( \big|\texttt C_{\texttt{\textsl f}}^{\texttt V}\big|^2
+ \big|\texttt C_{\texttt{\textsl f}}^{\texttt A}\big|^2
\Big) + 11\, \Big( \big|\texttt C_{\texttt{\textsl f}}^{\texttt S}\big|^2
+ \big|\texttt C_{\texttt{\textsl f}}^{\texttt P}\big|^2 \Big)
\vspace{2pt} \\ \,+~ 7.0\, \Big( \big|\tilde{\textsf c}_{\texttt{\textsl f}}^{\textsc v}\big|^2
+ \big|\tilde{\textsf c}_{\texttt{\textsl f}}^{\textsc a}\big|^2 \Big)
+ 2.3\, \Big( \big|\tilde{\textsf c}_{\texttt{\textsl f}}^{\textsc s}\big|^2
+ \big|\tilde{\textsf c}_{\texttt{\textsl f}}^{\textsc p}\big|^2 \Big) \Big] 10^7 \rm~GeV^4 \,,
\end{array} &
\nonumber \\
{\cal B}\big(\Sigma^+\to p\texttt{\textsl f}\bar{\texttt{\textsl f}}\big) & \,=
\begin{array}[t]{l} \! \Big[ 5.1\, \Big( \big|\texttt C_{\texttt{\textsl f}}^{\texttt V}\big|^2
+ \big|\texttt C_{\texttt{\textsl f}}^{\texttt A}\big|^2 \Big)
+ 26\, \big( |\texttt C_{\texttt{\textsl f}}^{\texttt S}|^2
+ |\texttt C_{\texttt{\textsl f}}^{\texttt P}|^2 \big)
\vspace{2pt} \\ \,+~ 1.8\, \Big( \big|\tilde{\textsf c}_{\texttt{\textsl f}}^{\textsc v}\big|^2
+ \big|\tilde{\textsf c}_{\texttt{\textsl f}}^{\textsc a}\big|^2 \Big)
+ 1.4\, \Big( \big|\tilde{\textsf c}_{\texttt{\textsl f}}^{\textsc s}\big|^2
+ \big|\tilde{\textsf c}_{\texttt{\textsl f}}^{\textsc p}\big|^2 \Big) \Big] 10^7 \rm~GeV^4 \,,
\end{array}
\nonumber \\
{\cal B}\big(\Xi^0\to\Lambda\texttt{\textsl f}\bar{\texttt{\textsl f}}\big) & \,=
\begin{array}[t]{l} \! \Big[ 9.3\, \Big( \big|\texttt C_{\texttt{\textsl f}}^{\texttt V}\big|^2
+ \big|\texttt C_{\texttt{\textsl f}}^{\texttt A}\big|^2 \Big)
+ 30\, \Big( \big|\texttt C_{\texttt{\textsl f}}^{\texttt S}\big|^2
+ \big|\texttt C_{\texttt{\textsl f}}^{\texttt P}\big|^2 \Big)
\vspace{2pt} \\ \,+~ 1.0\, \Big( \big|\tilde{\textsf c}_{\texttt{\textsl f}}^{\textsc v}|^2
+ \big|\tilde{\textsf c}_{\texttt{\textsl f}}^{\textsc a}\big|^2 \Big)
+ 0.4\, \Big( \big|\tilde{\textsf c}_{\texttt{\textsl f}}^{\textsc s}\big|^2
+ \big|\tilde{\textsf c}_{\texttt{\textsl f}}^{\textsc p}\big|^2 \Big) \Big] 10^7 \rm~GeV^4 \,,
\end{array}
\end{align}
\begin{align}
{\cal B}\big(\Xi^0\to\Sigma^0\texttt{\textsl f}\bar{\texttt{\textsl f}}\big) & \,=
\begin{array}[t]{l} \! \Big[ 0.29\, \Big( \big|\texttt C_{\texttt{\textsl f}}^{\texttt V}\big|^2
+ \big|\texttt C_{\texttt{\textsl f}}^{\texttt A}\big|^2 \Big)
+ 0.34\, \Big( \big|\texttt C_{\texttt{\textsl f}}^{\texttt S}\big|^2
+ \big|\texttt C_{\texttt{\textsl f}}^{\texttt P}\big|^2 \Big)
\vspace{2pt} \\ \,+~ 1.4\, \Big( \big|\tilde{\textsf c}_{\texttt{\textsl f}}^{\textsc v}\big|^2
+ \big|\tilde{\textsf c}_{\texttt{\textsl f}}^{\textsc a}\big|^2 \Big)
+ 0.20\, \Big( \big|\tilde{\textsf c}_{\texttt{\textsl f}}^{\textsc s}\big|^2
+ \big|\tilde{\textsf c}_{\texttt{\textsl f}}^{\textsc p}\big|^2 \Big) \Big] 10^7 \rm~GeV^4 \,,
\end{array}
\nonumber \\ \label{X2Sff}
{\cal B}\big(\Xi^-\to\Sigma^-\texttt{\textsl f}\bar{\texttt{\textsl f}}\big) & \,=
\begin{array}[t]{l} \! \Big[ 0.35\, \Big( \big|\texttt C_{\texttt{\textsl f}}^{\texttt V}\big|^2
+ \big|\texttt C_{\texttt{\textsl f}}^{\texttt A}\big|^2 \Big)
+ 0.42\, \Big( \big|\texttt C_{\texttt{\textsl f}}^{\texttt S}\big|^2
+ \big|\texttt C_{\texttt{\textsl f}}^{\texttt P}\big|^2 \Big)
\vspace{2pt} \\ \,+~ 1.7\, \Big( \big|\tilde{\textsf c}_{\texttt{\textsl f}}^{\textsc v}\big|^2
+ \big|\tilde{\textsf c}_{\texttt{\textsl f}}^{\textsc a}\big|^2 \Big)
+ 0.25\, \Big( \big|\tilde{\textsf c}_{\texttt{\textsl f}}^{\textsc s}\big|^2
+ \big|\tilde{\textsf c}_{\texttt{\textsl f}}^{\textsc p}\big|^2 \Big) \Big] 10^7 \rm~GeV^4 \,,
\end{array}
\end{align}
\begin{align} \label{BO2Xff}
{\cal B}\big(\Omega^-\to\Xi^-\texttt{\textsl f}\bar{\texttt{\textsl f}}\big) & \,=\, \Big[
7.9\, \Big( \big|\tilde{\textsf c}_{\texttt{\textsl f}}^{\textsc v}\big|\raisebox{1pt}{$^2$}
+ \big|\tilde{\textsf c}_{\texttt{\textsl f}}^{\textsc a}\big|^2 \Big)
+ 14\, \Big( \big|\tilde{\textsf c}_{\texttt{\textsl f}}^{\textsc s}\big|\raisebox{1pt}{$^2$}
+ \big|\tilde{\textsf c}_{\texttt{\textsl f}}^{\textsc p}\big|^2 \Big) \Big] 10^8 \rm~GeV^4 \,. &
\end{align}
All these results have incorporated the form factors mentioned in section \ref{hyperons}.
In the kaon sector, with \,$m_{\texttt{\textsl f}}^{}=0$,\, for the two-body decays we get
\begin{align} \label{BK2ff}
{\cal B}\big(K_L^{}\to\texttt{\textsl f}\bar{\texttt{\textsl f}}\big) & \,=\, 2.9\, \Big[
\big( {\rm Im}\, \tilde{\textsf c}_{\texttt{\textsl f}}^{\textsc s}\big)\raisebox{1pt}{$^2$}
+ \big( {\rm Re}\, \tilde{\textsf c}_{\texttt{\textsl f}}^{\textsc p} \big)\raisebox{1pt}{$^2$}
\Big] 10^{14} \rm~GeV^4 \,,
\nonumber \\
{\cal B}\big(K_S^{}\to\texttt{\textsl f}\bar{\texttt{\textsl f}}\big) & \,=\, 5.1\, \Big[
\big({\rm Re}\, \tilde{\textsf c}_{\texttt{\textsl f}}^{\textsc s}\big)\raisebox{1pt}{$^2$}
+ \big( {\rm Im}\, \tilde{\textsf c}_{\texttt{\textsl f}}^{\textsc p} \big)\raisebox{1pt}{$^2$}
\Big] 10^{11} \rm~GeV^4 \,,
\end{align}
leading to
\begin{align} \label{tcstcp}
\big|\tilde{\textsf c}_{\texttt{\textsl f}}^{\textsc s}\big|\raisebox{1pt}{$^2$}
+ \big|\tilde{\textsf c}_{\texttt{\textsl f}}^{\textsc p}\big|\raisebox{1pt}{$^2$}
& \,=\, \frac{3.4\times10^{-15}}{\rm GeV^4}\,
{\cal B}\big(K_L^{}\to\texttt{\textsl f}\bar{\texttt{\textsl f}}\big)
+ \frac{2.0\times10^{-12}}{\rm GeV^4}\,
{\cal B}\big(K_S^{}\to\texttt{\textsl f}\bar{\texttt{\textsl f}}\big) \,,
\end{align}
and for the four-body decays
\begin{align} \label{BK2pipiff}
{\cal B}\big(K^-\to\pi^-\pi^0\texttt{\textsl f}\bar{\texttt{\textsl f}}\big) & =
\Big[ 6.3 \Big( \big|\tilde{\textsf c}_{\texttt{\textsl f}}^{\textsc v}\big|\raisebox{1pt}{$^2$}
+ \big|\tilde{\textsf c}_{\texttt{\textsl f}}^{\textsc a}\big|\raisebox{1pt}{$^2$} \Big)
+ 2.0 \Big( \big|\tilde{\textsf c}_{\texttt{\textsl f}}^{\textsc s}\big|\raisebox{1pt}{$^2$}
+ \big|\tilde{\textsf c}_{\texttt{\textsl f}}^{\textsc p}\big|\raisebox{1pt}{$^2$} \Big) \Big]
10^5 \rm\,GeV^4 \,,
\nonumber \\
{\cal B}\big(K_L^{}\to\pi^0\pi^0\texttt{\textsl f}\bar{\texttt{\textsl f}}\big) & =
\Big[ 8.5 \big({\rm Re}\,\tilde{\textsf c}_{\texttt{\textsl f}}^{\textsc v}\big)\raisebox{1pt}{$^2$}
+ 8.5 \big({\rm Re}\,\tilde{\textsf c}_{\texttt{\textsl f}}^{\textsc a}\big)\raisebox{1pt}{$^2$}
+ 16 \big({\rm Im}\,\tilde{\textsf c}_{\texttt{\textsl f}}^{\textsc s}\big)\raisebox{1pt}{$^2$}
+ 16 \big({\rm Re}\,\tilde{\textsf c}_{\texttt{\textsl f}}^{\textsc p}\big)\raisebox{1pt}{$^2$}
\Big] 10^6 \rm\;GeV^4 \,.
\end{align}
Evidently, all the interference terms with different couplings have vanished as
\,$m_{\texttt{\textsl f}}^{}\to0$.\,

We can now look at a couple of representative instances with different choices of nonvanishing
couplings, which we take to be all real to ignore any new source of $CP$ violation.
If only $\tilde{\textsf c}_{\texttt{\textsl f}}^{\textsc s,\textsc p}$ are nonzero,
we find that the \,$K\to\slashed E$\, restrictions in eq.\,(\ref{K2inv}) are more
stringent than the \,$K\to\pi\pi'\slashed E$\, ones in eq.\,(\ref{K2ppinv}) and,
with the aid of eq.\,(\ref{tcstcp}), lead to
\,$\big|\tilde{\textsf c}_{\texttt{\textsl f}}^{\textsc s}\big|\raisebox{1pt}{$^2$}
+ \big|\tilde{\textsf c}_{\texttt{\textsl f}}^{\textsc p}\big|\raisebox{1pt}{$^2$}
< 2.2\times10^{-16}{\rm\;GeV}^{-4}$.\,
Combining this with eqs.\,\,(\ref{B2B'ff})-(\ref{BO2Xff}), we obtain
\begin{align}
{\cal B}\big(\Lambda\to n\texttt{\textsl f}\bar{\texttt{\textsl f}}\big) & \,<\,
5.0\times10^{-9} \,, &
{\cal B}\big(\Sigma^+\to p\texttt{\textsl f}\bar{\texttt{\textsl f}}\big) & \,<\,
3.0\times10^{-9} \,, &
\nonumber \\
{\cal B}\big(\Xi^0\to\Lambda\texttt{\textsl f}\bar{\texttt{\textsl f}}\big) & \,<\,
9.3\times10^{-10} \,, &
{\cal B}\big(\Omega^-\to\Xi^-\texttt{\textsl f}\bar{\texttt{\textsl f}}\big) & \,<\,
3.0\times10^{-7} \,, &
\end{align}
and smaller numbers for
${\cal B}\big(\Xi^{0,-}\to\Sigma^{0,-}\texttt{\textsl f}\bar{\texttt{\textsl f}}\big)$.
The upper ends of these ranges far exceed the respective SM values quoted in table \ref{smB2B'nunu}
but are still roughly two orders of magnitude beyond the expected BESIII reach.

If only $\tilde{\textsf c}_{\texttt{\textsl f}}^{\textsc v,\textsc a}$ are
nonzero, then according to eq.\,(\ref{BK2ff}) the \,$K\to\slashed E$\, constraints
no longer apply as  $\tilde{\textsf c}_{\texttt{\textsl f}}^{\textsc v,\textsc a}$
do not affect these decays in the \,$m_{\texttt{\textsl f}}^{}=0$\, limit,
but the \,$K\to\pi\pi'\slashed E$\, bounds still matter, the \,$K_L\to\pi^0\pi^0\slashed E$\,
one being the stronger and yielding
\,$\big({\rm Re}\,\tilde{\textsf c}_{\texttt{\textsl f}}^{\textsc v}\big)\raisebox{1pt}{$^2$}
+ \big({\rm Re}\,\tilde{\textsf c}_{\texttt{\textsl f}}^{\textsc a}\big)\raisebox{1pt}{$^2$}
<9.4\times10^{-14}{\rm\;GeV}^{-4}$.\,
This now translates into
\begin{align} \label{hypBRmax}
{\cal B}\big(\Lambda\to n\texttt{\textsl f}\bar{\texttt{\textsl f}}\big) & \,<\,
6.6\times10^{-6} \,, &
{\cal B}\big(\Sigma^+\to p\texttt{\textsl f}\bar{\texttt{\textsl f}}\big) & \,<\,
1.7\times10^{-6} \,, &
\nonumber \\
{\cal B}\big(\Xi^0\to\Lambda\texttt{\textsl f}\bar{\texttt{\textsl f}}\big) & \,<\,
9.4\times10^{-7} \,, &
{\cal B}\big(\Xi^0\to\Sigma^0\texttt{\textsl f}\bar{\texttt{\textsl f}}\big) & \,<\,
1.3\times10^{-6} \,, &
\nonumber \\
{\cal B}\big(\Xi^-\to\Sigma^-\texttt{\textsl f}\bar{\texttt{\textsl f}}\big) & \,<\,
1.6\times10^{-6} \,, &
{\cal B}\big(\Omega^-\to\Xi^-\texttt{\textsl f}\bar{\texttt{\textsl f}}\big) & \,<\,
7.5\times10^{-5} \,, &
\end{align}
most of which have upper values exceeding the corresponding estimated BESIII sensitivity
levels quoted in table \ref{smB2B'nunu}.
This suggests that BESIII might discover NP hints in these processes or, if not, come up with
improved restrictions on $\tilde{\textsf c}_{\texttt{\textsl f}}^{\textsc v,\textsc a}$.

If we let \,${\rm Im}\,\tilde{\textsf c}_{\texttt{\textsl f}}^{\textsc v,\textsc a}\neq0$,\,
bigger branching fractions than those in eq.\,(\ref{hypBRmax}) could be achieved with purely
imaginary $\tilde{\textsf c}_{\texttt{\textsl f}}^{\textsc v,\textsc a}$, as they would escape
the \,$K_L\to\pi^0\pi^0\slashed E$\, restraint and be subject only to the weaker
\,$K^-\to\pi^0\pi^-\slashed E$\, one, implying the mild limit
\,$\big({\rm Im}\,\tilde{\textsf c}_{\texttt{\textsl f}}^{\textsc v}\big)\raisebox{1pt}{$^2$}
+ \big({\rm Im}\,\tilde{\textsf c}_{\texttt{\textsl f}}^{\textsc a}\big)\raisebox{1pt}{$^2$}
< 6.4\times10^{-11}{\rm\,GeV}^{-4}$.\,
This serves to indicate further the benefit of measuring these hyperon decays, which may
test some of the NP couplings more stringently than the kaon decays.

\section{Conclusions\label{concl}}

We have explored the possibility that new physics contributes to the strangeness-changing
transition \,$s\to d\slashed E$,\, with missing energy in the final state.
Depending on the sizes of the NP couplings involved and the masses of the emitted invisible
particles, various changes could occur to the SM predictions for rare kaon and hyperon
\,$|\Delta S|=1$\, decays with missing energy.
We have learned that the current data on $K\to\pi\nu\bar\nu$ do not allow NP to influence
the hyperon decays considerably if the underlying operators have mostly parity-even quark portions.
On the other hand, if the NP operators have predominantly parity-odd, especially axial-vector,
quark parts, the restraints implied by the $K\to\rm invisible$ and $K\to\pi\pi'\nu\bar\nu$ data
are comparatively weaker.
We have demonstrated that NP with the latter kind of interactions could cause the hyperon
rates to be substantially amplified with respect to their SM expectations and have large
values potentially testable in the ongoing BESIII experiment.
This well illustrates that these rare hyperon decays and their kaon counterparts are
complementary to each other as probes of possible NP in \,$s\to d\slashed E$.\,

\acknowledgements

This research was supported in part by the MOE Academic Excellence Program
(Grant No. 105R891505).

\appendix

\section{Correspondences between quark and hadron transitions\label{quark-hadron}}

From the chiral Lagrangian that is at leading order in the derivative and $s$-quark-mass
($m_s$) expansions and describes the strong interactions among the lightest octet baryons
and mesons and decuplet baryons~\cite{Gasser:1983yg,Bijnens:1985kj,Jenkins:1991es},
one can extract correspondences between quark densities or currents and hadronic
transitions \cite{He:2005we}.
From the results of ref.\,\cite{He:2005we} pertaining to the \,$|\Delta S|=1$\, processes
under discussion, we can infer
\begin{align} \label{corresp}
\bar d\gamma_\eta^{}s & \;\Leftrightarrow\; -\sqrt{\frac{3}{2}}~\overline{n}\gamma_\eta^{}\Lambda
- \overline{p}\gamma_\eta^{}\Sigma^+
+ \sqrt{\frac{3}{2}}~ \overline{\Lambda}\gamma_\eta^{}\Xi^0
- \frac{1}{\sqrt2}\, \overline{\Sigma^0} \gamma_\eta^{}\Xi^0
+ \overline{\Sigma\bar{\hphantom{o}}}\gamma_\eta^{}\Xi^-
\nonumber \\ & ~~~ ~~ +\,
i \big( \pi^+\, \partial_\eta K^- - K^-\, \partial_\eta \pi^+ \big)
- \frac{i}{\sqrt2} \big( \pi^0\, \partial_\eta\, \overline{\!K}{}^0
-\, \overline{\!K}{}^0\, \partial_\eta \pi^0 \big)
+\, \cdots \,, & \hspace{5ex}
\end{align}
\begin{align}
\bar d s & \;\Leftrightarrow\;
\sqrt{\frac{3}{2}}~\frac{m_\Lambda^{}-m_N^{}}{\hat m-m_s^{}}\, \overline{n} \Lambda
+ \frac{m_\Sigma^{}-m_N^{}}{\hat m-m_s^{}}\, \overline{p}\,\Sigma^+
+ \sqrt{\frac{3}{2}}~ \frac{m_\Xi^{}-m_\Lambda^{}}{m_s^{}-\hat m}\, \overline{\Lambda}\,\Xi^0
\nonumber \\ & ~~~ ~~ +\,
\frac{m_\Xi^{}-m_\Sigma^{}}{\hat m-m_s^{}} \Bigg( \frac{\overline{\Sigma^0}\, \Xi^0}{\sqrt2}
- \overline{\Sigma\bar{\hphantom{o}}}\,\Xi^- \Bigg)
+\, B_0^{} \Bigg( \pi^+K^- - \frac{\pi^0\,\overline{\!K}{}^0}{\sqrt2} \Bigg) +\, \cdots \,, &
\end{align}
\begin{align} \label{dgg5s}
\bar d\gamma_\eta^{}\gamma_5^{}s & \;\Leftrightarrow\;
\frac{-D-3F}{\sqrt6}~ \overline{n}\gamma_\eta^{}\gamma_5^{}\Lambda
+ (D-F)\, \overline{p}\gamma_\eta^{}\gamma_5^{}\Sigma^+
- \frac{D-3F}{\sqrt6}~ \overline{\Lambda}\gamma_\eta^{}\gamma_5^{}\Xi^0   \hspace{7em}
\nonumber \\ & ~~~ ~~ -\,
\frac{D+F}{\sqrt2}~ \overline{\Sigma^0}\gamma_\eta^{}\gamma_5^{}\Xi^0 + (D+F)\,
\overline{\Sigma\bar{\hphantom{o}}}\gamma_\eta^{}\gamma_5^{}\Xi^-
\,+\, {\cal C}\, \overline{\Xi\bar{\hphantom{o}}}\, \Omega_\eta^-
\nonumber \\ & ~~~ ~~ +\,
\sqrt2\, f\, \partial_\eta^{}\, \overline{\!K}{}^0
+ \frac{\pi^+\partial_\eta\pi^0 - \pi^0\partial_\eta\pi^+}{f}\, K^-
- \frac{\pi^0\partial_\eta\, \overline{\!K}{}^0
-\, \overline{\!K}{}^0 \partial_\eta\pi^0}{3\sqrt2\, f}\, \pi^0
\,+\, \cdots \,,
\end{align}
\begin{align} \label{dg5s}
\bar d\gamma_5^{}s & \;\Leftrightarrow\; i\sqrt2\, B_0^{}\, f\, \overline{\!K}{}^0
- \frac{i B_0\, \pi^0\pi^0\, \overline{\!K}{}^0}{3\sqrt2\, f} \,+\, \cdots \,,   \hspace{15em}
\end{align}
where $m_{N,\Sigma,\Xi}^{}$ are isospin-averaged masses of the nucleons, $\Sigma^{\pm,0}$, and
$\Xi^{0,-}$, respectively, $\hat m$ is the average mass of the $u$ and $d$ quarks,
\,$B_0=m_K^2/(\hat m+m_s)$,\, with $m_K$ being the average mass of $K^0$ and $K^-$,
the free parameters $D$, $F$, and $\cal C$
occur in the lowest-order chiral Lagrangian and can be fixed from baryon decay data,
\,$f=f_K^{}/\sqrt2$,\, and the ellipses
stand for terms not relevant to our analysis.

At the same order in the chiral expansion, the baryonic matrix elements of
$\bar d\big(\gamma^\eta,1\big)\gamma_5^{}s$ and
$\langle\pi\pi'|\bar d\big(\gamma^\eta,1\big)\gamma_5^{}s|\bar K\rangle$
also receive contributions from kaon-pole diagrams involving
$\langle0|\bar d\big(\gamma^\eta,1\big)\gamma_5^{}s|\,\overline{\!K}{}^0\rangle$ from
eqs.\,\,(\ref{dgg5s}) and (\ref{dg5s}) and strong vertices from the lowest-order
strong chiral Lagrangian ${\cal L}_{\rm s}$.
The pertinent terms are given by
\begin{align}
{\cal L}_{\rm s}^{} \,\supset &~ \Bigg[
\frac{-D-3F}{\sqrt6}~ \overline{n} \gamma_\eta^{}\gamma_5^{}\Lambda
+ (D-F)\, \overline{p} \gamma_\eta^{}\gamma_5^{}\Sigma^+
- \frac{D-3F}{\sqrt6}~ \overline{\Lambda} \gamma_\eta^{}\gamma_5^{}\Xi^0
\nonumber \\ & ~~ -\,
\frac{D+F}{\sqrt2}~ \overline{\Sigma^0} \gamma_\eta^{}\gamma_5^{}\Xi^0
+ (D+F)\, \overline{\Sigma\bar{\hphantom{o}}} \gamma_\eta^{}\gamma_5^{}\Xi^-
+\, {\cal C}\, \overline{\Xi\bar{\hphantom{o}}}\, \Omega_\eta^-
\Bigg] \frac{\partial^\eta K^0}{\sqrt2\, f}
\nonumber \\ & \,+\,
\frac{\pi^0\mbox{\small$\stackrel{\scriptscriptstyle\leftrightarrow}{\partial}$}{}^\eta\pi^+}
{\sqrt8\,f^2}\,K^0\mbox{\small$\stackrel{\scriptscriptstyle\leftrightarrow}{\partial}$}_\eta^{}K^-
+ \frac{K^0\mbox{\small$\stackrel{\scriptscriptstyle\leftrightarrow}{\partial}$}{}^\eta\pi^0}
{12 f^2}\, \pi^0\mbox{\small$\stackrel{\scriptscriptstyle\leftrightarrow}{\partial}$}{}_\eta^{}\,
\overline{\!K}{}^0
+ \frac{m_K^2+m_\pi^2}{12 f^2}\, \pi^0 \pi^0 K^0\, \overline{\!K}{}^0 \,,
\end{align}
where
\,$X\raisebox{1pt}{\footnotesize$\stackrel{\scriptscriptstyle\leftrightarrow}{\partial}$}_\eta Y
= X_{\,}\partial_\eta Y-Y\partial_\eta X$.\,
From this and the previous paragraphs, we arrive at the hadronic matrix elements in eqs.\,\,(\ref{<B'B>}),
(\ref{<XO>}), and (\ref{<0K>})-(\ref{<K->pp>}), in the limit that \,$f_{+,0}^{}=1$.\,

In numerical work, we employ the observed hadron masses from ref.\,\cite{Tanabashi:2018oca} and
the light-quark mass values \,$\hat m=(m_u+m_d)/2=4.4$ MeV\, and \,$m_s=120$ MeV\,
at a renormalization scale of~1~GeV.
Moreover, we adopt \,$D=0.81$\, and \,$F=0.46$\, determined from fitting to the data on hyperon
semileptonic decays and \,${\cal C}=1.7$\, from the measurements of strong decays of the lightest
decuplet spin-3/2 baryons into an octet spin-1/2 baryon and a pion \cite{Tanabashi:2018oca}.

\section{Additional kaon decay formulas\label{Kformulas}}

The decay amplitudes for $K_{L,S}$ are connected to those for $K^0$ and
$\,\overline{\!K}{}^0$ through the approximate relations
\,$\sqrt2\, K_{L,S}=K^0\pm\, \overline{\!K}{}^0$.\,
Thus, for \,$K_{L,S}\to\texttt{\textsl f}\bar{\texttt{\textsl f}}$\, induced by
${\cal L}_{\texttt{\textsl f}}$ in eq.\,(\ref{Lnp}), with eq.\,(\ref{<0K>}) we derive
$S_{K\!_{L,S}\texttt{\textsl f}}$ and $P_{K\!_{L,S}\texttt{\textsl f}}$ in eq.\,(\ref{SK2ff}).

For \,$K^-\to\pi^-\texttt{\textsl f}\bar{\texttt{\textsl f}}$\, and
\,$K_L\to\pi^0\texttt{\textsl f}\bar{\texttt{\textsl f}}$,\, from eqs.\,\,(\ref{Lnp}) and
(\ref{<piK>}), we find the $S$ and $P$ terms in eq.\,(\ref{MK2pff}) to be
\begin{align}
S_{K^-\pi^-\texttt{\textsl f}}^{} & \,=\,
2 f_+^{}\, \slashed p_{\!K} \texttt C_{\texttt{\textsl f}}^{\texttt V}
+ B_0^{} f_0^{~} \texttt C_{\texttt{\textsl f}}^{\texttt S} \,,
\nonumber \\
P_{K^-\pi^-\texttt{\textsl f}}^{} & \,=\, 2 \Big[ (f_-^{}-f_+^{})m_{\texttt{\textsl f}}^{}
+ f_+^{}\,\slashed p_{\!K} \Big] \texttt C_{\texttt{\textsl f}}^{\texttt A}
+ B_0^{}f_0^{~}\texttt C_{\texttt{\textsl f}}^{\texttt P} \,,
\nonumber \\
S_{K_L^{}\pi^0\texttt{\textsl f}}^{} & \,=\,
-2i f_+^{}\, \slashed p_{\!K}\, {\rm Im}\, \texttt C_{\texttt{\textsl f}}^{\texttt V}
- B_0^{} f_0^{}\, {\rm Re}\, \texttt C_{\texttt{\textsl f}}^{\texttt S} \,,
\nonumber \\
P_{K_L^{}\pi^0\texttt{\textsl f}}^{} & \,=\,
2i \Big[ \big(f_+^{}-f_-^{}\big) m_{\texttt{\textsl f}}^{}
- f_+^{}\, \slashed p_{\!K} \Big] {\rm Im}\, \texttt C_{\texttt{\textsl f}}^{\texttt A}
- i B_0^{} f_0^{}\, {\rm Im}\, \texttt C_{\texttt{\textsl f}}^{\texttt P} \,.
\end{align}
For \,$K_S\to\pi^0\texttt{\textsl f}\bar{\texttt{\textsl f}}$,\, the $S$ and $P$ formulas
are equal to  $S_{K_L\pi^0\texttt{\textsl f}}$ and $P_{K_L\pi^0\texttt{\textsl f}}$  but with
\,${\rm Re}_{\!}\,\texttt C_{\texttt{\textsl f}}$\,  and
\,$-i{\rm Im}_{\!}\,\texttt C_{\texttt{\textsl f}}$\,  interchanged.
The differential rate of the $K_L$ decay is then
\begin{align} \label{G'KS}
\Gamma_{K_L^{}\to\pi^0\texttt{\textsl f}\bar{\texttt{\textsl f}}}' \,=\, \frac{\beta\lambda_{K^0\pi^0\,}^{1/2}f_0^2}
{128\pi^3 m_{K^0}^3} & \Bigg\{ \displaystyle \frac{3-\beta^2}{3f_0^2}
\lambda_{K^0\pi^0}^{}\,f_+^2\,
\big({\rm Im}_{\!}\,\texttt C_{\texttt{\textsl f}}^{\texttt V}\big)^{\!2}
+ \Bigg[ \frac{2\beta^2 f_+^2}{3 f_0^2} \lambda_{K^0\pi^0}^{}
+ \frac{4 m_{\texttt{\textsl f}}^2}{\hat s} \Delta_{K^0\pi^0}^4 \Bigg]
\big({\rm Im}_{\!}\,\texttt C_{\texttt{\textsl f}}^{\texttt A}\big)^{\!2}
\nonumber \\ & \;+\,
4 B_0^{} m_{\texttt{\textsl f}}^{} \Delta_{K^0\pi^0}^2\,
{\rm Im}_{\!}\,\texttt C_{\texttt{\textsl f}}^{\texttt A}\,
{\rm Im}_{\!}\,\texttt C_{\texttt{\textsl f}}^{\texttt P}
+ B_0^2 \Big[ \beta^2 \big({\rm Re}_{\!}\,\texttt C_{\texttt{\textsl f}}^{\texttt S}\big)^{\!2}
+ \big({\rm Im}_{\!}\,\texttt C_{\texttt{\textsl f}}^{\texttt P}\big)^2 \Big] \hat s \Bigg\} \,,
\end{align}
where \,$\Delta_{K\pi}^2=m_K^2-m_\pi^2$.\,
For $\Gamma_{K_S\to\pi^0\texttt{\textsl f}\bar{\texttt{\textsl f}}}'$, the expression is
obtainable from  $\Gamma_{K_L\to\pi^0\texttt{\textsl f}\bar{\texttt{\textsl f}}}'$  by
interchanging  ${\rm Re}_{\!}\,\texttt C_{\texttt{\textsl f}}$  and
${\rm Im}_{\!}\,\texttt C_{\texttt{\textsl f}}$,  while
$\Gamma_{K^-\to\pi^-\texttt{\textsl f}\bar{\texttt{\textsl f}}}'$ is the same as
\,$\Gamma_{K_L\to\pi^0\texttt{\textsl f}\bar{\texttt{\textsl f}}}'
+ \Gamma_{K_S\to\pi^0\texttt{\textsl f}\bar{\texttt{\textsl f}}}'$\,
but with $m_{K^0,\pi^0}$ replaced with $m_{K^-,\pi^-}$.

For the four-body decays, we have \cite{Kamenik:2011vy}
\begin{align} \label{G''K2ppff}
\frac{d^2\Gamma_{K^-\to\pi^0\pi^-\texttt{\textsl f}\bar{\texttt{\textsl f}}}}
{d\hat s\,d\hat\varsigma} \,=\, \frac{\beta_{\hat\varsigma\,}^{}
\tilde\lambda{}^{1/2\,}m_K^{}}{1536 \pi^5 f_K^2} & \Big[
\texttt F_1^{}\, \big|\tilde{\textsf c}_{\texttt{\textsl f}}^{\textsc v}\big|^2
+ \texttt F_1'\, \big|\tilde{\textsf c}_{\texttt{\textsl f}}^{\textsc a}\big|^2
+ \texttt F_2^{}\, \big|\tilde{\textsf c}_{\texttt{\textsl f}}^{\textsc s}\big|^2
+ \texttt F_2'\, \big|\tilde{\textsf c}_{\texttt{\textsl f}}^{\textsc p}\big|^2
\nonumber \\ & +\,
\texttt F_{12}^{} \big( {\rm Re}\,\tilde{\textsf c}_{\texttt{\textsl f}}^{\textsc a}\,
{\rm Re}\,\tilde{\textsf c}_{\texttt{\textsl f}}^{\textsc p} +
{\rm Im}\,\tilde{\textsf c}_{\texttt{\textsl f}}^{\textsc a}\,
{\rm Im}\,\tilde{\textsf c}_{\texttt{\textsl f}}^{\textsc p} \big) \Big] \,,
\end{align}
\begin{align}
\frac{d^2\Gamma_{K_L\to\pi^0\pi^0\texttt{\textsl f}\bar{\texttt{\textsl f}}}}
{d\hat s\,d\hat\varsigma} \,=\, \frac{\beta_{\hat\varsigma\,}^{} \tilde\lambda{}^{1/2\,}m_K^{}}
{3072 \pi^5 f_K^2} & \Big(
\texttt F_3^{}\, \big|{\rm Re}\,\tilde{\textsf c}_{\texttt{\textsl f}}^{\textsc v}\big|^2
+ \texttt F_3'\, \big|{\rm Re}\,\tilde{\textsf c}_{\texttt{\textsl f}}^{\textsc a}\big|^2
+\texttt F_4^{}\, \big|{\rm Im}\,\tilde{\textsf c}_{\texttt{\textsl f}}^{\textsc s}\big|^2
+ \texttt F_4'\, \big|{\rm Re}\,\tilde{\textsf c}_{\texttt{\textsl f}}^{\textsc p}\big|^2
\nonumber \\ & +\,
\texttt F_{34}^{}\, {\rm Re}\,\tilde{\textsf c}_{\texttt{\textsl f}}^{\textsc a}\,
{\rm Re}\,\tilde{\textsf c}_{\texttt{\textsl f}}^{\textsc p} \Big) \,,
\end{align}
where
\begin{align}
\beta_{\hat\varsigma}^{} & \,=\, \sqrt{1-\frac{4m_\pi^2}{\hat\varsigma}} \,, ~~~ ~~~
\hat\varsigma \,=\, (p_0^{}+p_-^{})^2 = (p_1^{}+p_2^{})^2 \,, ~~~ ~~~
\tilde\lambda \,=\, m_K^4-2\big(\hat s+\hat\varsigma\big)m_K^2
+ \big(\hat s-\hat\varsigma\big)\raisebox{1pt}{$^2$} \,,
\nonumber \\
\texttt F_{1,3}^{} & \,=\,
\lambda_{1,3}^{}\frac{2m_{\texttt{\textsl f}}^2+\hat s}{\hat s} \,, ~~~ ~~~~
\texttt F_{1,3}' \,=\,
\beta^2\lambda_{1,3}^{} + \lambda_{1,3}' \frac{m_{\texttt{\textsl f}}^2}{\hat s} \,, ~~~ ~~~~
\texttt F_{2,4}^{} \,=\, \beta^2 \texttt F_{2,4}' \,, ~~~ ~~~~
\texttt F_{2,4}' \,=\, \frac{\lambda_{1,3}' B_0^2 \hat s}{4 m_K^4} \,,
\nonumber \\
\texttt F_{12}^{} & \,=\,
-\lambda_1' \frac{B_0^{} m_{\texttt{\textsl f}}^{}}{m_K^2} \,, \hspace{8ex}
\texttt F_{34}^{} \,=\, -\lambda_3' \frac{B_0^{} m_{\texttt{\textsl f}}^{}}{m_K^2} \,,
\hspace{3em} \lambda_1^{} \,=\, \frac{\beta_{\hat\varsigma}^2}{m_K^4}
\bigg(\hat s\hat\varsigma+\frac{\tilde\lambda}{12}\bigg) , ~~~ ~~~~
\lambda_3^{} \,=\, \frac{\tilde\lambda}{4 m_K^4} \,,
\nonumber \\
\lambda_1' & \,=\, \frac{\beta_{\hat\varsigma}^2
\tilde\lambda}{2\big(m_K^2-\hat s\big)\raisebox{1pt}{$^2$}} \,, ~~~~ ~~~~~
\lambda_3' \,=\,
\frac{12\hat s\hat\varsigma+3\tilde\lambda}{2\big(m_K^2-\hat s\big)\raisebox{1pt}{$^2$}} \,.
\end{align}
We compute the rates by integrating the double-differential rates in eq.\,(\ref{G''K2ppff})
over the intervals \,$4m_{\texttt{\textsl f}}^2\le\hat s\le(m_K-2m_\pi)^2$\, and
\,$4m_\pi^2\le\hat\varsigma\le\big(m_K^{}-\hat s^{1/2}\big)\raisebox{1pt}{$^2$}$.\,



\end{document}